\documentclass[11pt]{article}
\usepackage{amsmath,amsfonts,latexsym,graphicx,amssymb}
\usepackage{amsfonts}
\usepackage{bm}
\date{}

\newcommand{\ot}{{\,\otimes\,}}
\newcommand{{\Cd}}{{\mathbb{C}^d}}

\def\oper{{\mathchoice{\rm 1\mskip-4mu l}{\rm 1\mskip-4mu l}%
{\rm 1\mskip-4.5mu l}{\rm 1\mskip-5mu l}}}
\def\<{\langle}
\def\>{\rangle}
\newtheorem{theorem}{Theorem}

\newtheorem{definition}{Definition}
\numberwithin{equation}{section}

\begin{document}
\title{\bf Remarks on the GNS Representation and the Geometry of Quantum States} \author{Dariusz
Chru\'sci\'nski  \\
Institute of Physics, Nicolaus Copernicus University,\\
Grudzi\c{a}dzka 5/7, 87--100 Toru\'n, Poland \\ \\
Giuseppe Marmo\\
Dipartamento di Scienze Fisiche, Universit\'a ``Federico II" di
Napoli \\
and Instituto Nazionale di Fisica Nucleare, Sezione di Napoli, \\
Complesso Universitario di Monte Sant Angelo,\\
Via Cintia, I-80126 Napoli, Italy }

\maketitle

\begin{abstract}

It is shown how to introduce a geometric description of the
algebraic approach to the non-relativistic quantum mechanics. It
turns out that the GNS representation provides not only symplectic
but also Hermitian realization of a `quantum Poisson algebra'. We
discuss alternative Hamiltonian structures emerging out of different
GNS representations which provide a natural setting for quantum
bi-Hamiltonian systems.

\end{abstract}

\maketitle

\section{Introduction}
\setcounter{equation}{0}

The important role played by geometry in the formulation of
theories aimed at the descriptions of fundamental interactions
cannot be denied. At the moment classical theories like mechanics,
electromagnetism, Einstein's General Relativity, Yang-Mills gauge
theories and thermodynamics have reached a very high degree of
geometrization. The same cannot be said for quantum theories, even
though the relevance of geometric structures, like the symplectic
structure, may be traced back to Segal and Mackey
\cite{Mackey,Segal}, and since quite few papers have been written
on the subject [3--16].

For historical reasons \cite{BOOK} the geometrical structures are
hidden in the standard algebraic setting of quantum mechanics
(notably the Dirac formulation) because one starts from the Hilbert
space and identifies the space of physical states with the
associated complex projective space, which in a natural way calls
for a differential geometric treatment
\cite{Kibble,Cantoni1,Cirelli,K-bundle}, however, for simplicity,
computations are carried out on the initial Hilbert space. In this
approach the $\mathbb{C}^*$-algebra, which contains observables as
real elements, arises as a derived concept
--- as complex valued functions on the complex projective space
endowed with  an appropriate associative product even though
non-commutative and non-local.

In this short note we would like to consider, on the contrary, a
different approach, often called an algebraic one, where the Hilbert
space loses its primary importance. The primary object one starts
with is an abstract $\mathbb{C}^*$-algebra containing an algebra of
quantum observables and the Hilbert space is a secondary concept
which may be derived by constructing particular representation of
$\mathcal{A}$ in the spirit of GNS construction (see e.g.
\cite{Bratteli,Kadison}). Algebraic approach started with the work
of Haag and Kastler \cite{Haag-Kastler} and then it was used mainly
in the mathematical approach to quantum field theory \cite{Haag}.
This approach is much more flexible than the standard one: a Hilbert
space is not a priori given but it is derived by using a given state
of the system.  Different states give rise to different realization
of the original algebra as an algebra of operators, that is, one is
able to derive different Hilbert spaces, inner products and
multiplication rules in the space of operators acting in the
constructed Hilbert space.

The analogy that we pursue is the following: in many classical
situations one is presented with a Poisson manifold and looks for
a symplectic realization of its Poisson algebra. Here, in a
similar way, we would like to consider the `quantum Poisson
algebra' of the complexification of the space of observables and
search for a Hermitian realization of it. We observe that this
`Hermitian realization' is the essence of the well known GNS
construction.

We start with a $\mathbb{C}^\star$-algebra $\mathcal{A}$. It may be
decomposed into two real vector spaces of real and imaginary
elements
\[  \mathcal{A}_{\rm re} = \{ a+a^*\ | \ a \in \mathcal{A}\}\ , \ \ \ \
\mathcal{A}_{\rm im} = \{ a-a^*\ | \ a \in \mathcal{A}\}\ ,
\]
respectively. There is a one-to-one correspondence between
$\mathcal{A}_{\rm re}$ and $\mathcal{A}_{\rm im}$ by means of
multiplication by `$i$'. Consider the space of density states
$\mathcal{D}(\mathcal{A})$ over $\mathcal{A}$ which is a convex body
spanned by extremal (pure) states $\mathcal{D}^1(\mathcal{A})$. Out
of the vector space $\cal A$ we may construct the  dual  space
$L(\mathcal{A})$ by taking real combinations of
$\mathcal{D}^1(\mathcal{A})$, then we may immerse $\mathcal{A}_{\rm
re}$ into the space of linear functionals on the real vector space
$L(\mathcal{A})$. Now, using the commutator product in $\mathcal{A}$
the linear subspace $\mathcal{A}_{\rm re}$ induces a Poisson
structure on $\mathcal{D}(\mathcal{A})$. It shows that the real
vector space $L(\mathcal{A})$ constructed out of
$\mathcal{D}^1(\mathcal{A})$ may be thought of as the dual to
$\mathcal{A}_{\rm re}$ (or equivalently to $\mathcal{A}_{\rm im}$).
$L(\mathcal{A})$ may be endowed with a Poisson structure and gives
rise to the Lie algebra of Hamiltonian vector fields. It turns out
that  Hamiltonian vector fields associated with linear maps on
$L(\mathcal{A})$ -- i.e. element from
$\mbox{Lin}(L(\mathcal{A}),\mathbb{R})$ -- may be thought of as
derivations of the product available on $\mathcal{A}$ or of the
pointwise  product that we may construct on the polynomials of
$\mbox{Lin}(L(\mathcal{A}),\mathbb{R})$. The pointwise (commutative)
product identifies the Poisson bracket as those of a Poisson algebra
(commutative algebra on which derivations act). Moreover, one may
introduce a noncommutative $\star$-product in
$\mbox{Lin}(L(\mathcal{A}),\mathbb{C})$, that is, in the
$\mathbb{C}^*$-algebra of complex valued function on the space of
states $L(\mathcal{A}) \supset \mathcal{D}(\mathcal{A})$. In this
way the Poisson bracket `$f\star g - g \star f$' with $f,g \in
\mbox{Lin}(L(\mathcal{A}),\mathbb{C})$, should be considered as a
quantum bracket in the sense of Dirac. Summarizing: the Poisson
bracket on the dual space to $\mathcal{A}_{\rm re}$ may be used to
generate derivations for the commutative algebra of polynomials and
therefore as a `classical' Poisson algebra. The same Poisson bracket
when restricted to linear functions defines derivations for the
usual (noncommutative) operator product defined in the space of
operators but thought of as functions on the dual space. This gives
rise to a `quantum' Poisson algebra.

Having noticed that on the space of self-adjoint elements of a
$C^*$-algebra one has a Lie algebra structure and a Jordan structure
, one may ``geometrize'', i.e. describe these products in terms of
tensorial objects, by using functions and tensors defined on the
dual of the $C^*$-algebra (Lie algebra). In this way we obtain a
Poisson manifold and a Lie-Jordan product associated with a Jordan
tensor.

The main idea for our geometrization uses the dual space of the
$C^*$-algebra. We recall that in the works of Gelfand and
collaborators the dual space of Banach algebras the study of the
dual spaces has been found extremely useful. In the sixties Fell
\cite{Fell} has considered dual spaces of $C^*$-algebras and Banach
algebras providing many interesting results. Here we would like to
take a different route and, to present the geometrical ideas more
clearly, we restrict to finite dimensional algebras. We shall use,
however, an intrinsic formulation, paving the way to an extension to
the infinite dimensional situation. We shall use coordinates only to
allow the reader to became more familiar with our construction.

The paper is organized as follows: we start with a short review of
the geometric formulation of the standard non-relativistic Quantum
Mechanics. Then we review GNS construction  and provide its simple
illustration in the case of matrix algebra in section \ref{GNS-IL}.
Section \ref{SYMP} shows how the GNS construction gives rise to the
symplectic realization of the Poisson algebra of observables: either
via a corresponding Hilbert space defining the representation space
of GNS or via the associated complex projective space. In section
\ref{Bi} we discuss alternative Hamiltonian structures emerging out
of different GNS realizations of the original
$\mathbb{C}^*$-algebra. It turns out that the GNS representation
provides a natural arena for quantum bi-Hamiltonian systems. Final
conclusions are collected in the last section.

\section{Geometric formulation of Quantum Mechanics}

We first review very briefly the geometrical formulation of
Quantum Mechanics starting  with a standard Hilbert space
formulation. The essential steps are the following: The
probabilistic interpretation requires that the physical carrier
space of our formulation should be identified with the space of
rays
\[  \mathbb{C}_0 \ \longrightarrow\ \mathcal{H}_0\
\longrightarrow\ \mathcal{R}(\mathcal{H})\ ,\] where $\mathbb{C}_0 =
\mathbb{C}- 0$ and $\mathcal{H}_0 = \mathcal{H} - \{0\}$. The {\it
true} space of quantum states -- space of rays
$\mathcal{R}(\mathcal{H})$ -- is nothing but the complex projective
Hilbert space $\mathbb{P}\mathcal{H}$. Now, to replace vectors and
linear transformations by tensor fields we have to replace
$\mathcal{H}$ with $T\mathcal{H}$, its tangent bundle, which may be
identified as a Cartesian product $T\mathcal{H} \sim \mathcal{H}
\times \mathcal{H}$. Any vector $\varphi \in {\cal H}$ gives rise to
a vector field $X_\varphi : {\cal H} \longrightarrow T{\cal H}$
defined by
\begin{equation}\label{}
    X_\varphi(\psi) := (\psi,\varphi) \in {\cal H}\times {\cal H}
    \ .
\end{equation}
Similarly an endomorphism $A : {\cal H} \longrightarrow {\cal H}$
gives rise to a map $T_A : T{\cal H} \longrightarrow T{\cal H}$
defined as follows
\begin{equation}\label{}
    T_A(\psi,\varphi):= (\psi,A\varphi)\ .
\end{equation}
Moreover, one introduces a complex structure ${\cal J} : T{\cal H}
\longrightarrow T{\cal H}$ defined by the $(1,1)$-tensor field
\begin{equation}\label{}
{\cal J}(\psi,\varphi):= (\psi,i\varphi)\ ,
\end{equation}
and a linear structure $\Delta : {\cal H} \longrightarrow T{\cal H}$
defined by the Liouville vector field
\begin{equation}\label{}
\Delta(\psi):= (\psi,\psi)\ ,
\end{equation}
and finally the so called phase--vector field $\Gamma : {\cal H}
\longrightarrow T{\cal H}$ defined by $\Gamma := {\cal J} \circ
\Delta\ $, i.e.
\begin{equation}\label{}
\Gamma(\psi) =(\psi,i\psi) \ .
\end{equation}
 In this way the Hermitian product $\< \psi|\varphi\>$
on $\mathcal{H}$ is replaced by an Hermitian tensor field
\begin{equation}\label{}
    {\cal K}(X_{\varphi_1},X_{\varphi_2})(\psi) :=
    \<\varphi_1|\varphi_2\>\ .
\end{equation}
 On the corresponding real differential manifold $\mathcal{H}^{\rm R}$
 the real part of ${\cal K}$ is a Riemannian metric tensor `$g$' while its imaginary
part is a symplectic tensor field `$\omega$'
\begin{equation}\label{}
    \mathcal{K} = g + i\omega\ ,
\end{equation}
together with
\begin{equation}\label{}
    \omega(X,Y) = g(\mathcal{J}X,Y)\ .
\end{equation}
The above tensor fields endow $\mathcal{H}$  with the structure of
a K\"ahler manifold. When written in a contravariant form $G$ and
$\Lambda$, respectively,  give rise to two bi-differential
operators which may be used to define two brackets on the space of
one-forms. We should notice that the symmetric tensor may also be
associated with a second order differential operator (Laplacian).
These tensor fields may be used to define the metric structure and
Poisson bracket on the space of rays $\mathcal{R}(\mathcal{H})$.
Note, however, that neither $G$ nor $\Lambda$ can  be directly
projected from $\mathcal{H}$ to $\mathcal{R}(\mathcal{H})$. It is
easy to show that tensor fields which are  projectable  are given
by
\begin{equation}\label{}
    \widetilde{G} := e^{\sigma} G - \Delta \ot
    \Delta - \Gamma \ot \Gamma\ ,
\end{equation}
and
\begin{equation}\label{}
    \widetilde{\Lambda} := e^\sigma \Lambda - (\Delta \ot
    \Gamma - \Gamma \ot \Delta) \ ,
\end{equation}
where the conformal factor $e^\sigma \geq 0$ is defined by
$\sigma(\psi) := \ln \<\psi|\psi\>$. Now, projected tensor fields
allow for the definition of two products in the space of functions
on $\mathcal{R}(\mathcal{H})$: the symmetric bracket
\begin{equation}\label{}
\{f_1,f_2\}_+ := \widetilde{G}(df_1,df_2) + f_1\cdot f_2 \ ,
\end{equation}
and antisymmetric Poisson bracket
\begin{equation}\label{}
\{f_1,f_2\} := \widetilde{\Lambda}(df_1,df_2) \ .
\end{equation}
The above operations are defined for arbitrary real valued
functions from $\mathcal{F}(\mathcal{R}(\mathcal{H}))$. In this
formulation quantum observables are defined to be  functions from
$\mathcal{F}(\mathcal{R}(\mathcal{H}))$ whose Hamiltonian vector
fields are at the same time also Killing vector fields, i.e.
\begin{equation}\label{}
\mathcal{F}_{\rm K}(\mathcal{R}(\mathcal{H})) := \{\ f\in
\mathcal{F}(\mathcal{R}(\mathcal{H})) \ |\ {\bf L}_{X_f}
\widetilde{G} = 0 \ \}\ ,
\end{equation}
where $X_f = \widetilde{\Lambda}(df)$. We call such `$f$' a {\it
K\"ahlerian function}. To deal with complex valued functions, we
need the extension from real valued functions to complex valued
functions. A complex valued function is K\"ahlerian iff both real
and imaginary parts are K\"ahlerian.  On this selected space of
K\"ahler functions we may define an associative bilinear product $f
\star g$ corresponding to the Hermitian tensor $\widetilde{\cal K} =
\widetilde{G} + i\widetilde{\Lambda}$:
\begin{equation}\label{}
f\star g := f \cdot g + \frac 12\, \widetilde{\cal K}(df,dg) \ .
\end{equation}
One shows that for any two K\"ahler functions `$f$' and `$g$' the
nonlocal product `$f \star g$' defines a K\"ahler function.
Consider now  the complexified space
$\mathcal{F}^{\mathbb{C}}_{\rm K}(\mathcal{R}(\mathcal{H}))$. Let
us observe that any complex valued K\"ahlerian function  on
$\mathcal{R}(\mathcal{H})$ corresponds to an operator $A \in
\mathcal{B}(\mathcal{H})$
\begin{equation}\label{}
    A \ \longrightarrow\ f_A([\psi]) :=
    \frac{\<\psi|A\psi\>}{\<\psi|\psi\>}\ ,
\end{equation}
that is, $f_A$ is an expectation value function. It is easy to
show that
\begin{equation}\label{}
    f_A \star f_B = f_{AB}\ .
\end{equation}
Quantum observables correspond to real valued K\"ahlerian functions
and hence they are represented by Hermitian operators on $\cal H$.
The complexified space $\mathcal{F}^{\mathbb{C}}_{\rm
K}(\mathcal{R}(\mathcal{H}))$ equipped with the above noncommutative
$\star$-product provides a realization of a $\mathbb{C}^*$-algebra
corresponding to a $\mathbb{C}^*$-algebra of bounded operators
acting on the initial Hilbert space $\mathcal{H}$, i.e. the algebra
$\mathcal{B}(\mathcal{H})$ \cite{K-bundle}.

Consider now a general K\"ahlerian manifold
$(\mathcal{M},\widetilde{\cal K})$ not necessarily a projective
Hilbert space $\mathbb{P}\mathcal{H}$.  It is clear that one may
define a nonlocal $\star$-product
\begin{equation}\label{}
f\star g := f \cdot g + \frac 12\, \widetilde{\cal K}(df,dg) \ ,
\end{equation}
for arbitrary $f,g \in \mathcal{F}_{\rm
K}^{\mathbb{C}}(\mathcal{M})$. Now, for arbitrary $\cal M$ the
corresponding space of complex valued K\"ahlerian functions is not
closed under $\star$--product. The Poisson bracket
\begin{equation}\label{}
    \{f,g\} = \frac i2 (f \star g - g \star f) \ ,
\end{equation}
does belong to $\mathcal{F}_{\rm K}^{\mathbb{C}}(\mathcal{M})$,
however, the symmetric bracket
\begin{equation}\label{}
    \{f,g\}_+ = \frac 12 (f \star g + g \star f) \ ,
\end{equation}
in general is not a K\"ahlerian function. The condition that the
space of K\"ahlerian function over $\cal M$ is closed with respect
to symmetric bracket puts strong conditions on the K\"ahler
structure. It turns out that it is equivalent to the very intricate
geometric property of $\cal M$, namely, that  holomorphic sectional
curvature of $\cal M$ is constant \cite{CMP}. This in turn implies
that $\cal M$ is a projective Hilbert space $\mathbb{P}\mathcal{H}$
or the covering space of the symplectic orbit in $u^*(\mathcal{H})$.
Thus only orbits of the unitary group are associated with
$\mathbb{C}^*$-algebras -- they will be given by the generating
functions of the Hamiltonian action of the unitary group.

After the GNS construction one should be able to prove that
realization of the $\mathbb{C}^*$-algebras are in one-to-one
correspondence with the action of the unitary group on the
K\"ahler manifold.

\section{Finite dimensional setting}

Let us illustrate the above geometrical formulation for finite
dimensional Hilbert space ${\cal H} = \mathbb{C}^{n+1}$. Denote by
$\{|e_j\>\}$, with $j=0,1,\ldots,n$, an orthonormal basis in
$\mathbb{C}^{n+1}$. Then for any vector $\psi \in \mathbb{C}^{n+1}$
one has
\begin{equation}\label{}
    \<e_j|\psi\> = z^j =: q^j + ip^j \ ,
\end{equation}
and
\begin{equation}\label{}
    |d\psi\> = dz^j|e_j\> = (d q^j + idp^j)|e_j\> \ .
\end{equation}
Using Cartesian coordinate system $(q^j,p^k)$ on ${\cal H}^{\rm
R}$ one easily finds
\begin{equation}\label{}
    \Delta = q^j\frac{\partial}{\partial q^j} + p^j \frac{\partial}{\partial
    p^j} \ ,
\end{equation}
and
\begin{equation}\label{}
    \Gamma = p^j\frac{\partial}{\partial q^j} - q^j \frac{\partial}{\partial
    p^j} \ .
\end{equation}
Moreover,  the Hermitian tensor field reads as follows
\begin{eqnarray}\label{}
    \mathcal{K}(d\psi,d\psi) &=& \overline{(d q^j + idp^j)} \ot (d q^k +
    idp^k)\<e_j|e_k\> \nonumber \\ &=& (d q^j \ot dq^k + d p^j
    \ot dp^k)\<e_j|e_k\>  \\ &+& i(d q^j \ot dp^k - d p^j
    \ot dq^k)\<e_j|e_k\>\ \nonumber .
\end{eqnarray}
The corresponding contravariant tensors  $G$ and $\Lambda$ are
therefore given by
\begin{equation}\label{}
G = \left( \frac{\partial}{\partial q_j} \ot
\frac{\partial}{\partial q_k}  + \frac{\partial}{\partial p_j} \ot
\frac{\partial}{\partial p_k} \right) \<e_j|e_k\> \ ,
\end{equation}
and for the Poisson tensor
\begin{equation}\label{}
\Lambda = \left( \frac{\partial}{\partial p_j} \ot
\frac{\partial}{\partial q_k}  - \frac{\partial}{\partial p_k} \ot
\frac{\partial}{\partial q_j} \right) \<e_j|e_k\> \ .
\end{equation}
Finally, one may introduce the following local coordinates on
$\mathbb{C}P^n \equiv \mathbb{P}\mathcal{H}$:
\begin{equation}\label{}
    w_k = \frac{z_k}{z_0} \ ,
\end{equation}
for $z_0\neq 0$. Using projective coordinates $(w_1,\ldots,w_n)$
one obtains the following formula for $\widetilde{\mathcal{K}}$
\begin{equation}\label{}
    \widetilde{\mathcal{K}} = \sum_{i,j=1}^n \frac{(1+
    |w|^2)\delta_{ij} - \overline{w}_i w_j}{(1 + |w|^2)^2} \, dw_i
    \ot d\overline{w}_j \ ,
\end{equation}
where $|w|^2 = \sum_{k=1}^n w_k \overline{w}_k$.

Interestingly, K\"ahlerian functions on complex projective space are
eigenfunctions of the corresponding Laplacian $\Delta_n$. It is well
known \cite{Ikeda} that the spectrum of the Laplacian on
$\mathbb{C}P^n$ is given by\footnote{Actually, in \cite{Ikeda} the
eigenvalues differ by a factor `4'. It corresponds to different
normalization of $\Delta_n$. Our convention reproduces quantum
mechanical result `$l(l+1)$' on $\mathbb{C}P^1 \cong S^2$.}
\begin{equation}\label{}
    \lambda_{n,l} = - l(n+l)\ , \ \ \ \ l=0,1,2,\ldots\ ,
\end{equation}
and the corresponding multiplicity of $\lambda_{n,l}$ reads as
follows \cite{Boucetta}
\begin{equation}\label{}
    N_{n,l} = n(n+2l)\left(\frac{(n+l-1)!}{n!\,l!} \right)^2\ .
\end{equation}
Note that for $l=1$ one obtains
\begin{equation}\label{}
    N_{n,1} = n(n+2) = (n+1)^2-1\ ,
\end{equation}
that is, it reproduces the dimension of the space of traceless
Hermitian operators in $\mathbb{C}^{n+1}$. Now, one may prove that
$f$ is K\"ahlerian iff
\begin{equation}\label{}
    \Delta_n f = 0 \ , \ \ \ \mbox{or} \ \ \ \Delta_n f = \lambda_{n,1} f \
    ,
\end{equation}
 that is, $f$ is either a zero mode of
$\Delta_n$,  or it is an eigenvector of $\Delta_n$ corresponding to
the first nonvanishing eigenvalue `$-(n+1)$'.  Since zero mode span
1-dimensional space one finds that the space of K\"ahlerian
functions is $(n+1)^2$--dimensional, i.e. has the same dimension as
the space of Hermitian operators in $\mathbb{C}^{n+1}$.

To see how it works let us consider the simplest case $n = 1$. The
corresponding projective space $\mathbb{C}P^1$  is given by the
Bloch sphere $S^2$ and the eigenvalue problem $\Delta_1 f =
\lambda_{1,1} f$ is well known from the theory of angular momentum.
One has
\begin{equation}\label{Ylm}
\Delta_1  Y_{lm} = -l(l + 1)Y_{lm}\ ,
\end{equation}
where $Y_{lm}$ are spherical harmonics and the integer $m$ runs
from $-l$ to $l$. Note, that (\ref{Ylm}) implies that $l = 0$ or
$l = 1$. In the first case $Y_{00}$ defines a constant function on
$S^2$, whereas in the second case we have three independent dipole
functions $Y_{11} = x$, $Y_{1-1} = y$, and $Y_{10} = z$.  A
constant function corresponds  to the identity operator
$\mathbb{I}$. One easily checks that dipole functions correspond
to Pauli matrices: $\sigma_x$, $\sigma_y$, and $\sigma_z$.

\section{Review of the GNS construction}
\setcounter{equation}{0}

The geometrization we have presented starts from the Hilbert space
formulation of Quantum Mechanics. Now, we would like to consider
directly the $\mathbb{C}^*$-algebra approach and provide a direct
geometrization of this approach. According to the algebraic approach
to quantum theory \cite{Haag-Kastler,Haag,B-Segal} the basic notion
is the space of observables which consists of real elements of a
$\mathbb{C}^*$-algebra with unity $\mathcal{A}$. Note, that
observables carry a structure of Jordan algebra equipped with the
symmetric Jordan product
\begin{equation}\label{}
    a \circ b := \frac 12 (ab + ba)\ ,
\end{equation}
and of Lie algebra with the antisymmetric Lie product
\begin{equation}\label{}
    [a ,b] := \frac i2 (ab - ba)\ .
\end{equation}
These two products recover an original product in $\cal A$:
\begin{equation}\label{}
    ab = a \circ b - i [a,b]\ .
\end{equation}

In this approach states are represented by positive, normalized
linear functionals on $\mathcal{A}$,  that is $\omega \in {\cal
D}(\mathcal{A})$ (set of states over $\mathcal{A}$) if for any $a
\in \mathcal{A}$ one has $\omega(aa^*)\geq 0$ and $\omega(\oper)
=1$, where $\oper$ stands for a unit element in $\mathcal{A}$. That
is, the set of states ${\cal D}(\mathcal{A})$ may be embedded
$\mathcal{D}(\mathcal{A}) \hookrightarrow L(\mathcal{A})$ into the
dual of ${\cal A}$.

The Hilbert space which in the traditional Schr\"odinger formalism
is considered as a primary object does not any longer play this
distinguished role. In the algebraic approach it appears as a
secondary object which is constructed out of a selected state of the
system under consideration. The construction which associates with
each state $\omega$ over $\mathcal{A}$ a particular Hilbert space
$\mathcal{H}_\omega$ is known as the GNS-construction: note that
$\omega$ defines the following pairing between elements from
$\mathcal{A}$
\begin{equation}\label{}
    \< a | b\>_\omega = \omega(a^*b)\ .
\end{equation}
Positivity of $\omega$ guarantees that $\< a|a\>_\omega \geq 0$ but
this pairing may be degenerate, that is, one may have $\<
a|a\>_\omega = 0$ for $a \neq 0$. To cure this problem one
introduces the so called Gelfand  ideal $\mathcal{J}_\omega$
consisting of all elements $a \in \mathcal{A}$ such that
$\omega(a^*a)=0$. The set of classes
$\mathcal{A}/\mathcal{J}_\omega$ defines a pre-Hilbert space and the
positive definite scalar product on $\mathcal{A}/\mathcal{J}_\omega$
\begin{equation}\label{scalar}
    \< \Psi_a|\Psi_b\> = \omega(a^*b)\ ,
\end{equation}
where $\Psi_a$ and $\Psi_b$ stand for the equivalence classes of
$a$ and $b$, respectively:
\begin{equation}\label{}
    \Psi_a = [ a + \mathcal{J}_\omega ] \ , \ \ \ \  \Psi_b = [ b + \mathcal{J}_\omega ] \
    .
\end{equation}
Formula (\ref{scalar}) does not depend on the choice of elements
$a$ and $b$ from the
 classes $\Psi_a$ and $\Psi_b$. Finally, completing
 $\mathcal{A}/\mathcal{J}_\omega$ in the norm topology induced by the
 scalar product (\ref{scalar}) one obtains a Hilbert space
 $\mathcal{H}_\omega$. This construction gives rise to the
 following representation of $\mathcal{A}$: for any $a \in
 \mathcal{A}$ one defines a linear operator $\pi_\omega(a)$ acting
 on $\mathcal{H}_\omega$ as follows
 \begin{equation}\label{}
    \pi_\omega(a)\Psi_b = \Psi_{ab} \ ,
\end{equation}
where $b$ is any element from the  class $\Psi_b$. Moreover, if
$\pi_\omega$ is a faithful representation (that is, $a\neq 0
\Longrightarrow \pi_\omega(a) \neq 0$) then the operator norm of
$\pi_\omega(a)$ equals the $\mathbb{C}^*$-norm of $a$ in
$\mathcal{A}$. It is clear that the GNS-construction provides a
cyclic representation with a cyclic vector $\Omega \in
\mathcal{H}_\omega$ corresponding to the class of the unit element
in $\mathcal{A}$, i.e. $\Omega = \Psi_\oper$. Moreover,
\begin{equation}\label{}
    \omega(a) = \< \Omega| \pi_\omega(a) | \Omega\> \ .
\end{equation}
By the duality $\cal A$ acts on ${\cal D}({\cal A})$ and hence a
Hilbert space corresponding to a state $\omega \in {\cal D}({\cal
A})$ is nothing but an orbit of $\cal A$ passing through $\omega$,
i.e. $\mathcal{H}_\omega \equiv \mathcal{A} \cdot \omega$.

Note that given any element $b \in \mathcal{A}$ one obtains a new
vector $\Psi=\pi_\omega(b)\Omega \in\mathcal{H}_\omega$. If $\Psi$
has norm one, this defines a new state $\omega_\Psi$ over
$\mathcal{A}$ given by
\begin{equation}\label{vector}
\omega_\Psi(a) = \< \Psi| \pi_\omega(a) | \Psi\> \ ,
\end{equation}
or equivalently
\begin{equation}\label{}
    \omega_\Psi(a) = \omega(b^*ab)\ ,
\end{equation}
for all $a \in \mathcal{A}$. One calls states over $\mathcal{A}$
defined by (\ref{vector}) vector states of representation
$\pi_\omega$. More general states may be defined by density
operators $\rho$ in $\mathcal{B}(\mathcal{H}_\omega)$ via
\begin{equation}\label{rho-states}
    \omega_\rho(a) = \mbox{Tr}( \rho\,\pi_\omega(a))\ .
\end{equation}
One calls all states (\ref{rho-states}) a folium of the
representation $\pi_\omega$. Let us recall that two
representations $\pi_1$ and $\pi_2$ of $\mathcal{A}$ defined on
two Hilbert spaces $\mathcal{H}_1$ and $\mathcal{H}_2$,
respectively, are equivalent, if there exists a unitary
intertwiner $U : \mathcal{H}_1 \longrightarrow \mathcal{H}_2$ such
that
\begin{equation}\label{}
    U\pi_1(a)U^* = \pi_2(a)\ ,
\end{equation}
for any $a \in \mathcal{A}$. The GNS representation is universal
in the following sense: if $\pi$ is a cyclic  representation of
$\mathcal{A}$ defined on $\mathcal{H}$, then the vector
representation $\omega_\Psi$ defined via (\ref{vector}) is
equivalent to $\pi$ for any normalized $\Psi \in {\cal H}_\omega$.

 Now, a state $\omega$ over
$\mathcal{A}$ is pure if and only if it cannot be written as a
convex combination of other states from $\mathcal{D}(\mathcal{A})$.
It is clear that the set of pure states (denoted by
$\mathcal{D}^1(\mathcal{A})$)  defines a set of extremal points of
the convex body $\mathcal{D}(\mathcal{A})$. The importance of pure
states follows from the following

\begin{theorem} \label{IRREP}
A GNS representation $\pi_\omega$ of $\mathcal{A}$ is irreducible
if and only if $\omega$ is a pure state over $\mathcal{A}$.
\end{theorem}

\section{Illustration: GNS for matrix algebra} \label{GNS-IL}
\setcounter{equation}{0}

To illustrate how the Hilbert space emerges out of a
$\mathbb{C}^*$-algebra $\mathcal{A}$ consider the following simple
example. Let $\mathcal{A} = \mathcal{B}(\mathbb{C}^n)$, i.e. the
algebra of $n \times n$ complex matrices. Any semi-positive
operator $\omega \in \mathcal{B}(\mathbb{C}^n)$ defines a state
over $\mathcal{A}$ via
\begin{equation}\label{}
\omega(A) = \mbox{Tr}(\omega A)\ ,
\end{equation}
for $A \in \mathcal{A}$.  Now, for any $A,B \in \mathcal{A}$ one
defines the inner product
\begin{equation}\label{inner}
    \<A|B\>_\omega = \omega(A^*B) = \mbox{Tr}(B\omega A^*)  \ .
\end{equation}
Let $\omega$ be a rank-1 projector. Then there is a basis
$\{e_k\}$ in $\mathbb{C}^n$ such that $\omega = |e_1\>\<e_1|$.
Hence
\begin{equation}\label{}
 \<A|B\>_\omega = \sum_{k=1}^n
     \overline{A}_{k1} B_{k1} =: \sum_{k=1}^n
     \overline{a}_{k} b_{k}\ ,
\end{equation}
with $a_k := A_{k1}$ and $b_k:=B_{k1}$. Note, that the
corresponding Gelfand ideal is defined as follows:
\begin{equation}\label{}
    \mathcal{J}_\omega = \{ \, X \in \mathcal{A}\ | \ X_{k1}=0\ ,
    k=1,\ldots,n\, \}\ ,
\end{equation}
that is,
\begin{equation}\label{}
\<A+X|B+Y\>_\omega = \<A|B\>_\omega\ ,
\end{equation}
for any $X,Y\in \mathcal{J}_\omega$.
 This shows that the Hilbert space $\mathcal{H}_\omega \equiv {\cal A}/\mathcal{J}_\omega \subset {\cal A}^*$
emerging out of rank-1 projector is nothing but $\mathbb{C}^n$. It
is, therefore, clear that the GNS representation of $\mathcal{A}$
in $\mathcal{H}_\omega$ reproduces the defining representation of
$\mathcal{B}(\mathbb{C}^n)$. To see that the Hilbert space does
indeed depend upon the state over $\mathcal{A}$ consider rank-$m$
projector in $\mathcal{B}(\mathbb{C}^n)$ given by $\omega = p_1
|e_1\>\<e_1| + \ldots +  p_m |e_m\>\<e_m|$, with $p_1,\ldots,p_m
>0$ and $p_1 + \ldots + p_m =1$. One obtains
\begin{equation}\label{AB-m}
 \<A|B\>_\omega = \sum_{k=1}^n
     \left(p_1 \overline{A}_{k1} B_{k1} + \ldots +  p_m \overline{A}_{km} B_{km}
     \right) =: \sum_{k=1}^n \left( \overline{a}^{(1)}_{k} b^{(1)}_{k} + \ldots
     + \overline{a}^{(m)}_{k} b^{(m)}_{k} \right) \ ,
\end{equation}
where
\begin{equation}\label{}
{a}^{(j)}_{k} = \sqrt{p_j}\, A_{kj} \ , \ \ \ \ \ {b}^{(j)}_{k} =
\sqrt{p_j}\, B_{kj}\ .
\end{equation}
The r.h.s. of (\ref{AB-m}) may be called ``normal form'' of the
Hermitian product. This construction shows  very clearly that the
Hermitian product on the Hilbert space we have constructed depends
on the state. In a sense the ``preparation'' of the state $\omega$
selects the Hermitian structure in ${\cal H}_\omega$.

Note that the corresponding Gelfand ideal is defined as follows:
\begin{equation}\label{}
    \mathcal{J}_\omega = \{ \, X \in \mathcal{A}\ | \ X_{kj}=0\ ,
    k=1,\ldots,n\ , \ j=1,\ldots,m\, \}\ .
\end{equation}
If $m=n$, then $\mathcal{J}_\omega$ is trivial.  It shows that the
resulting Hilbert space reads as $ \mathcal{H}_\omega \cong
\mathbb{C}^n \oplus \ldots \oplus \mathbb{C}^n$ ($m$ copies). Now,
the corresponding GNS representation $\pi_\omega$ is no longer
irreducible in $\mathbb{C}^n \oplus \ldots \oplus \mathbb{C}^n$
but decomposes into the direct sum of $m$ irreducible (defining)
representations
\begin{equation}\label{}
    \pi_\omega = \bigoplus_{k=1}^m \pi_k \ ,
\end{equation}
that is $\pi_\omega(A) = \mathbb{I}_m \ot A$, where $\mathbb{I}_m$
is an $m \times m$ identity matrix.

Let us observe that the form of the inner product (\ref{inner})
suggests to define a new multiplication rule in the space of
operators in $\mathcal{B}(\mathbb{C}^n)$, indeed from
\begin{equation}\label{inner-new}
    \<A|B\>_\omega =  \mbox{Tr}(B\omega A^*)  \ ,
\end{equation}
we may set
\begin{equation}\label{}
    A\cdot_\omega B := A\omega B \ .
\end{equation}
It defines a new associative product in
$\mathcal{B}(\mathbb{C}^n)$. As we shall see in section \ref{Bi}
this new product turns out to be very useful to define
bi-Hamiltonian structure for quantum evolution
\cite{Dubrovin,GS1,CGM,EIMM-a,EIMM}.

\section{Geometrization of algebraic structures}

Let $V$ be a vector space and consider its dual $V^*$. One may
imbed $V$ into its bi-dual $(V^*)^*$
\begin{equation}\label{}
    V \ni v \ \longrightarrow \ \widehat{v} \in {\cal F}(V^*) \ ,
\end{equation}
by
\begin{equation}\label{}
    \widehat{v}(\alpha) := \alpha(v)\ ,
\end{equation}
for $\alpha \in V$.  This imbedding allows to deal with polynomial
functions directly associated with multilinear functions on $V^*
\times \ldots \times V^*$ by restricting them to the diagonal, i.e.
for any multilinear function
\begin{equation}\label{}
    f \ : \ V^*\times \ldots \times V^*\ \longrightarrow\
    \mathbb{R}\ ,
\end{equation}
its reduction $\widetilde{f}(\alpha) := f(\alpha,\ldots,\alpha)$ is
a polynomial function in ${\cal F}(V^*)$. Note, that for any
$v_1,v_2 \in V$ one defines the product $\widehat{v}_1 \cdot
\widehat{v}_2$ by
\begin{equation}\label{V-product}
(\widehat{v}_1 \cdot \widehat{v}_2)(\alpha) :=
\widehat{v}_1(\alpha) \cdot \widehat{v}_2(\alpha)\ ,
\end{equation}
with $\alpha \in V^*$. Clearly, $\widehat{v}_1 \cdot \widehat{v}_2$
defines a polynomial in ${\cal F}(V^*)$.

Suppose now that $V$ carries an additional structure defined by a
bilinear operation
\begin{equation}\label{}
    B \ : \ V \times V \ \longrightarrow \ V\ .
\end{equation}
Let us observe that we may use $B$ to define a 2-tensor field
$\tau_B$ by setting
\begin{equation}\label{product-2}
    \tau_B(d\widehat{v}_1 ,d\widehat{v}_2)(\alpha) :=
    \alpha(B(v_1,v_2))\ .
\end{equation}
Using
\begin{equation}\label{}
    d(\widehat{v}_1 \cdot \widehat{v}_2) = (d\widehat{v}_1
    )\cdot \widehat{v}_2 + \widehat{v}_1\cdot (d\widehat{v}_2)\ ,
\end{equation}
one finds
\begin{equation}\label{}
\tau_B(d\widehat{v} ,d(\widehat{v}_1\cdot \widehat{v}_2))=
\tau_B(d\widehat{v} ,d\widehat{v}_1)\cdot \widehat{v}_2 +
\widehat{v}_1\cdot \tau_B(d\widehat{v},d \widehat{v}_2)\ ,
\end{equation}
which shows that $\tau_B(d\widehat{v})$ is a derivation of the
product (\ref{V-product}). In this sense we may speak of the
geometrical description of the binary product by introducing  the
tensor field $\tau_B$ which defines a bi-differential operator.

Now, we shall consider the special cases when $B$ endows $V$ with
the structure of Lie algebra or Jordan algebra. Let us start with a
Lie algebra $g=(V,B)$, where $B$ is skew-symmetric and  satisfies
the Jacobi identity
\begin{equation}\label{}
    B(v_1,B(v_2,v_3)) + {\rm cyclic\ permutations} = 0 \ .
\end{equation}
It is evident that $\Lambda:=\tau_B$ defines a Poisson tensor on
$\mathcal{F}(V^*)$. Moreover, one may prove that in this case
$\Lambda(d\widehat{v})$ is also a derivation of (\ref{product-2}).

\vspace{.3cm}

\noindent {\it Example:} as an example consider the 3-dimensional
Lie algebra $V=\mathbb{R}^3$ defined by
\begin{equation}\label{}
    B(v_1,v_2)=a_3 v_3\ , \ \ B(v_2,v_3)=a_1 v_1\ , \ \ B(v_3,v_1)=a_2 v_2\ ,
\end{equation}
with $a_1,a_2,a_3 \in \mathbb{R}$. Defining 3 coordinate functions
\begin{equation}\label{}
    x_1 = \widehat{v}_1\ , \ \  x_2 = \widehat{v}_2\ , \ \  x_3 = \widehat{v}_3\ ,
\end{equation}
together with
\begin{equation}\label{}
    \mathcal{C}(x_1,x_2,x_3) = \frac 12 (a_1 x_1^2 + a_2 x_2^2 +
    a_3 x_3^2) \ ,
\end{equation}
one finds for the Poisson tensor
\begin{equation}\label{}
 \Lambda = \epsilon_{ijk}\, \frac{\partial \mathcal{C}}{\partial
    x_i}\, \frac{\partial }{\partial
    x_j} \wedge \frac{\partial }{\partial  x_k}\ .
\end{equation}
Note, that $\cal C$ is a Casimir function, i.e. $\Lambda({\cal
C},f)=0$. By properly choosing $a_1,a_2,a_3$ one obtains all
unimodular 3-dimensional Lie algebras.  \hfill $\Box$

Consider now $V$ equipped with a Jordan product
\begin{equation}\label{product-3}
    B(v_1,v_2) = v_1 \circ v_2\ .
\end{equation}
The corresponding Riemann tensor ${\cal R}:= \tau_B$ is defined by
\begin{equation}\label{}
    {\cal R}(d\widehat{v}_1,d\widehat{v}_2)(\alpha) = \alpha(v_1
    \circ v_2)\ .
\end{equation}
Now, contrary to the Poisson tensor, ${\cal R}(d\widehat{v})$ is a
derivation of (\ref{V-product}) but no longer a derivation of the
Jordan product (\ref{product-3}).

Finally, let $(V,\cdot)$ be a $\mathbb{C}^*$-algebra. It is
equipped both with the antisymmetric Lie product
\begin{equation}\label{}
    B(v_1,v_2) := \frac i2 (v_1 \cdot v_2 - v_2 \cdot v_1) \ ,
\end{equation}
and the symmetric Jordan product
\begin{equation}\label{}
    B'(v_1,v_2) := \frac 12 (v_1 \cdot v_2 + v_2 \cdot v_1) \ .
\end{equation}
Let $\Lambda := \tau_B$ and ${\cal R}:= \tau_{B'}$ be the
corresponding Poisson and Riemann tensors. Note, that these two
structures endow the real elements of $\mathbb{C}^*$-algebra with a
structure of a Lie-Jordan algebra

\begin{definition}
A Lie-Jordan algebra $(\mathcal{B},\circ,[\ ,\ ])$ is a real vector
space endowed with two bilinear operations  `$\circ$' and $[\ ,\ ]$
with the following properties
\begin{eqnarray*}\label{}
a \circ b &=& b \circ a\ , \\
{}[a,b] &=& - [b,a] \ .
\end{eqnarray*}
Moreover, Lie-Jordan brackets satisfy the Leibniz rule
\begin{equation}\label{}
    [a,b\circ c]  = [a,b]\circ c + b\circ [a,c]\ ,
\end{equation}
and Jacobi identity
\begin{equation}\label{}
    [a,[b,c]] = [[a,b],c] + [b,[a,c]]\ .
\end{equation}
Finally,
\begin{equation}\label{}
    (a\circ b)\circ c - a\circ (b \circ c) = \lambda^2 [[a,c],b] \
    ,
\end{equation}
for some real number $\lambda$.
\end{definition}
Hamiltonian vector fields on $V^*$ constructed with $\Lambda$ define
derivation for the Jordan product. This construction completes the
`geometrization' of a $\mathbb{C}^*$-algebra.

\vspace{.3cm}

\noindent {\it Example:} Consider the Lie algebra $u(2)$ in the
defining representation on $\mathbb{C}^2$. It is spanned by 4
anti-Hermitian matrices $v_\alpha = i \sigma_\alpha$, with
$\alpha=0,1,2,3$, where
\begin{equation}\label{}
    \sigma_0 = \left( \begin{array}{cc} 1 & 0 \\ 0 & 1 \end{array} \right)
    \ , \ \ \
    \sigma_1 = \left( \begin{array}{cc} 0 & 1 \\ 1 & 0 \end{array}  \right)
    \ , \ \ \
    \sigma_2 = \left( \begin{array}{cc} 0 & -i \\ i & 0 \end{array} \right)
    \ , \ \ \
    \sigma_3 = \left( \begin{array}{cc} 1 & 0 \\ 0 & -1 \end{array} \right) \ ,
\end{equation}
are Pauli matrices. Now, let us define coordinate functions
\begin{equation}\label{}
    y_\alpha(A) = \frac 12 \, \mbox{Tr}(\sigma_\alpha A) \ ,
\end{equation}
for $A \in u(2)$. Using the well known property
\begin{equation}\label{}
    \sigma_k \sigma_l = i \epsilon_{klm} \sigma_m \ ,
\end{equation}
one obtains the following formulae for the Poisson tensor
\begin{equation}\label{}
    \Lambda = 2 \sum_{k,l,m=1}^3\, \epsilon_{klm} \, y_k \, \frac{\partial}{\partial
    y_l} \wedge \frac{\partial}{\partial y_m} \ ,
\end{equation}
and for the Riemann tensor
\begin{equation}\label{}
    {\cal R} = \frac{\partial}{\partial y_0} \ot_s \sum_{k=1}^3
    y_k\, \frac{\partial}{\partial y_k} + y_0 \sum_{k=1}^3\ \frac{\partial}{\partial
    y_k} \ot \frac{\partial}{\partial y_k}\ ,
\end{equation}
where $\ot_s$ stands for the symmetrized tensor product, i.e.
$a\ot_s b = a\ot b + b \ot a$.

Moreover, the Hamiltonian vector fields $H_\alpha$ corresponding
to coordinate functions $y_\alpha$, i.e. $H_\alpha =
\Lambda(y_\alpha,\cdot)$,  are defined as follows
\begin{equation}\label{}
    H_0 = 0 \ , \ \ \ H_k = \sum_{l,m=1}^3\, \epsilon_{klm} \, y_m \, \frac{\partial}{\partial
    y_l}\ ,
\end{equation}
for $k=1,2,3$. Finally the gradient vector fields $X_\alpha$ defined
by $X_\alpha := {\cal R}(y_\alpha,\cdot)$ read as follows
\begin{equation}\label{}
    X_0 =   \sum_{\alpha=0}^3 y_\alpha \frac{\partial}{\partial
    y_\alpha}\ , \ \ \   \  X_k =  y_k \frac{\partial}{\partial
    y_0} + y_0 \frac{\partial}{\partial y_k} \ ,
\end{equation}
for $k=1,2,3$. Note, that
\begin{equation}\label{}
    [X_\alpha,X_\beta] = y_\alpha \frac{\partial}{\partial y_\beta} - y_\beta  \frac{\partial}{\partial
    y_\alpha}\ .
\end{equation}
Finally, one may show that the union of these two distributions
$H_k$ and $X_k$ $(k=1,2,3$) generates $SL(2,{\mathbb{C}})$.

\section{Hermitian and K\"ahlerian realizations via GNS construction} \label{SYMP}
\setcounter{equation}{0}

Each pure state $\omega$ over $\mathcal{A}$ gives rise to
irreducible representation $\pi_\omega$ of $\mathcal{A}$ in the
Hilbert space $\mathcal{H}_\omega$. It is clear that real elements
in $\mathcal{A}$ are represented via $\pi_\omega$ by self-adjoint
operators in $\mathcal{B}(\mathcal{H}_\omega)$ which are in a
one-to-one correspondence with the real Lie algebra
$u(\mathcal{H}_\omega)$ of the unitary group
$U(\mathcal{H}_\omega)$. The symplectic action of
$U(\mathcal{H}_\omega)$ on $\mathcal{H}_\omega$ by $(U,\Psi)
\longrightarrow U\Psi$, provides us with the corresponding momentum
map
\begin{equation}\label{mu-H}
    \mu_\omega\ :\ \mathcal{H}_\omega \ \longrightarrow\
    u^*(\mathcal{H}_\omega) \ ,
\end{equation}
where $u^*(\mathcal{H}_\omega)$ denotes the dual of the Lie algebra
$u(\mathcal{H}_\omega)$. The map is defined by
\begin{equation}\label{}
    \mu_\omega(\psi) = |\psi\>\<\psi| \ .
\end{equation}
Note, that $u^*(\mathcal{H}_\omega)$ is a Poisson manifold and hence
(\ref{mu-H}) provides a symplectic realization. Let us recall that a
symplectic realization of a Poisson manifold $(M,\Lambda)$ is a
Poisson map $\Phi : S \longrightarrow M$, where $(S,\Omega)$ is a
symplectic manifold. When $S$ is a symplectic vector space one calls
$\Phi$ a classical Jordan-Schwinger map \cite{J-S}. When $S$ is a
Hilbert space we shall call it {\it Hermitian realization}.

Now, the action of $U({\cal H}_\omega)$ on ${\cal H}_\omega$ induces
the symplectic action of $U(\mathcal{H}_\omega)$ on the space of
rays $\mathcal{R}(\mathcal{H}_\omega)$ via
\begin{equation}\label{}
    (U,[\psi]) \ \longrightarrow [U\psi] \ .
\end{equation}
The above action provides us with the corresponding momentum map
\begin{equation}\label{mu-RH}
    \widetilde{\mu}_\omega\ :\ \mathcal{R}(\mathcal{H}_\omega) \ \longrightarrow\
    u^*(\mathcal{H}_\omega) \ ,
\end{equation}
defined by
\begin{equation}\label{}
    \widetilde{\mu}_\omega([\psi]) = \frac{\mu_\omega(\psi)}{\<\psi|\psi\>}\ .
\end{equation}
Now, because the above action preserves also the Riemann tensor, the
momentum map relates also this tensor on
$\mathcal{R}(\mathcal{H}_\omega)$ with the symmetric tensor on
$u^*(\mathcal{H}_\omega)$ obtained from the Jordan algebra on
$u^*(\mathcal{H}_\omega)$. The Hermitian tensor on
$\mathcal{R}(\mathcal{H}_\omega)$ will be therefore
$\mu_\omega$--related to a corresponding tensor on
$u^*(\mathcal{H}_\omega)$. Again (\ref{mu-RH}) provides a symplectic
realization. We shall call a symplectic realization $\Phi : S
\longrightarrow M$ {\it K\"ahlerian realization} if $S$ is a
submanifold of the complex  projective  space. Actually, it was
proved by Gromov \cite{Gr1,Gr2} (see also \cite{Gr3}) that any
compact K\"ahlerian manifold may be immersed into the complex
projective space (in the analogy to the Whitney theorem about
embedding of a manifold into the Euclidean space $\mathbb{R}^N$).

The linear structure of $u^*(\mathcal{H}_\omega)$ allows for convex
combinations in $\mu_\omega(\mathcal{R}(\mathcal{H}_\omega))\subset
u^*(\mathcal{H}_\omega)$ and hence enables one to consider density
operators. Consider now a general mixed state $\varphi$ over
$\mathcal{A}$. The corresponding GNS-representation $\pi_\varphi$ is
no longer irreducible on $\mathcal{H}_\varphi$. One has therefore
the following direct sum decomposition
\begin{equation}\label{}
    \pi_\varphi\, =\, \bigoplus_\alpha\, \pi_\alpha \ ,
\end{equation}
where $\pi_\alpha$ are irreducible representations of $\mathcal{A}$
on $\mathcal{H}_\alpha$, and
\begin{equation}\label{}
    \mathcal{H}_\varphi\,  =\, \bigoplus_\alpha\,
    \mathcal{H}_\alpha\ .
\end{equation}
It implies that a `vacuum' vector $\Omega \in \mathcal{H}_\varphi$
decomposes as follows
\begin{equation}\label{}
    \Omega\, =\, \bigoplus_\alpha\,\Omega_\alpha\ , \ \ \ \
    \Omega_\alpha \in \mathcal{H}_\alpha\ .
\end{equation}
It is clear that each irreducible representation $\pi_\alpha$
corresponds to a pure state $\varphi_\alpha$ defined by
\begin{equation}\label{}
    \varphi_\alpha(a) = \frac{1}{p_\alpha}\, \< \Omega_\alpha|\pi_\alpha(a)|\Omega_\alpha\>_\alpha \
    ,
\end{equation}
where $\<\ |\ \>_\alpha$ denotes the scalar product in
$\mathcal{H}_\alpha$, and
\begin{equation}\label{p-alpha}
    p_\alpha = \< \Omega_\alpha|\Omega_\alpha\>_\alpha\ .
\end{equation}
Normalization of $\Omega$ implies $\sum_\alpha p_\alpha=1$. It shows
that a mixed state $\varphi$ decomposes as the following convex
combination of pure states $\varphi_\alpha$
\begin{equation}\label{}
    \varphi = \sum_\alpha\,p_\alpha \,\varphi_\alpha\ ,
\end{equation}
that is
\begin{equation}\label{}
    \varphi(a) = \sum_\alpha\, \< \Omega_\alpha|\pi_\alpha(a)|\Omega_\alpha\>_\alpha\
    .
\end{equation}

\section{Alternative Hamiltonian structures} \label{Bi}
\setcounter{equation}{0}

We stress that different states  over $\mathcal{B}(\mathbb{C}^n)$
give rise to different GNS representations and hence to different
realizations of the Hilbert spaces. As we already observed a state
over $\mathcal{B}(\mathbb{C}^n)$ corresponds to a positive $n \times
n$ matrix $K$ (we replaced abstract $\omega$ by $K$) and hence may
be used to define an alternative scalar product in $\mathbb{C}^n$
\begin{equation}\label{}
    z\cdot_K w = \sum_{k,l=1}^n \overline{z}_k K_{kl} w_l\ ,
\end{equation}
for any $z,w \in \mathbb{C}^n$. One recovers the standard form if
$K=\mathbb{I}$, that is
\begin{equation}\label{}
    z\cdot w = \sum_{k=1}^n \overline{z}_k w_k\ .
\end{equation}
Different inner products in $\mathcal{H}$ are associated with
different multiplication rules in the space of operators
\begin{equation}\label{}
    A \cdot_K B = A\cdot K \cdot B\ ,
\end{equation}
for any $A,B \in \mathcal{B}(\mathbb{C}^n)$. Note that the product
`$\,\cdot_K\,$' defined by the above formula is associative, and
hence $(\mathcal{B}(\mathbb{C}^n),\cdot_K)$ carries a structure of
a $\mathbb{C}^*$-algebra.

With these alternative associative products we may associate
alternative Lie algebra structures and alternative Jordan
algebras. According to what we have said earlier they are similar
to the alternative Poisson structures we find in classical
dynamics when dealing with bi-Hamiltonian systems and complete
integrability. To carry the analogy consider now quantum dynamics
governed by the Hamiltonian $H$ and suppose, that
\begin{equation}\label{}
    [H,K] = H\cdot K - K \cdot H = 0 \ .
\end{equation}
Note that
\begin{equation}\label{}
    [A,H] = A\cdot H - H \cdot A = A\cdot_K H_K - H_K \cdot_K A =:
    [A,H_K]_K\ ,
\end{equation}
with
\begin{equation}\label{}
    H_K = K^{-1} \cdot H\ .
\end{equation}
 It proves that one has two alternative descriptions of quantum evolution: either
the standard Heisenberg equation
\begin{equation}\label{}
    i\hbar\dot{A} = [A,H]\ ,
\end{equation}
or the equivalent description using deformed multiplication
\begin{equation}\label{}
    i\hbar\dot{A} = [A,H_K]_K\ .
\end{equation}

Consider now a description of the quantum systems in terms of the
Wigner-Weyl formalism. In this approach an operator $A$ on $\cal
H$ is represented by a function $f_A$ on a classical phase space
$\cal P$. The commutative product in the space of functions ${\cal
F}({\cal P})$ is deformed into the noncommutative $\star$-product
such that $f_A \star f_B = f_{AB}$.  Moreover,  in the classical
limit
\begin{equation}\label{}
    \lim_{\hbar \rightarrow 0} \frac 1\hbar \{\!\{ f,g \}\!\}_{\star } =
 \{f,g\}  \ ,
\end{equation}
where $\{\!\{f,g\}\!\}_\star = \frac{1}{2i}(f\star g - g \star
f)$. As was already found by Rubio \cite{Rubio}, any associative
local product in the commutative algebra of functions ${\cal
F}({\cal P})$ has the following form
\begin{equation}\label{}
    f\cdot_k g := fkg\ ,
\end{equation}
where $f,k,g \in \mathcal{F}(\mathcal{P})$ and $k > 0$. Therefore,
one may use this new product `$\cdot_k$' to define an alternative
$\star_k$-product
\begin{equation}\label{}
    f_A \star_k f_B := f_A \star k \star f_B\ ,\ .
\end{equation}
It  gives rise to the following equation of motion
\begin{equation}\label{}
    i\hbar\dot{f}_{A} = \{\!\{ f_A,f_H \}\!\}_{\star k} \ ,
\end{equation}
where the Moyal-like $\star_k$ bracket reads as follows
\begin{equation}\label{}
\{\!\{ f_A,f_B \}\!\}_{\star k} = \frac{1}{2i} (f_A \star_k f_B -
f_B \star_k f_A)\ .
\end{equation}
Note, that in the `classical limit'
\begin{equation}\label{}
\lim_{\hbar\rightarrow 0} \frac 1\hbar \{\!\{ f_A,f_B
\}\!\}_{\star k} = k\, \{f_A,f_B\} + f_A X_k(f_B) - f_BX_k(f_A)\ ,
\end{equation}
where $X_k$ is a Hamiltonian vector field corresponding to $k$.
Interestingly, the `classical limit' of the Moyal $\star_k$
bracket is not a Poisson one but a Jacobi bracket. For $k=1$ one
has $X_k =0$ and hence one recovers the standard Poisson bracket.
Similarly, the `classical limit' of the symmetric Jordan bracket
gives
\begin{equation}\label{}
\lim_{\hbar\rightarrow 0}\ \frac 12 ( f_A \star_k f_B + f_B
\star_k f_A) = f_A \cdot_k f_B\ .
\end{equation}
It shows that there are alternative deformation quantization schemes
depending upon the associative product $f\cdot_k g$ in the original
commutative algebra $\mathcal{F}(\mathcal{P})$. The additional
function `$k$' has been related to the Kubo-Martin-Schwinger (KMS)
state \cite{Sternheimer,St2}.

\section{Conclusions}
\setcounter{equation}{0}

The contribution of this paper is to start directly from
$\mathbb{C}^*$-algebra to `geometrize' it and then use the GNS
construction to recover the Hilbert space. As a matter of fact in
our geometric version we naturally obtain a K\"ahler bundle
defined on the space of states. Let us recall
\begin{definition}
A K\"ahler bundle is a triple $(P,B,p)$, where $P$ (total space)
and $B$ (base) are topological spaces and $p : P \longrightarrow
B$ is a surjective continuous map. Moreover, for each $b \in B$
the fiber $p^{-1}(b)$ is a K\"ahler manifold.
\end{definition}
Indeed, the space of states over $\mathbb{C}^*$-algebra $\cal A$ is
naturally embedded into the dual $L({\cal A})$
\begin{equation}\label{e}
    e \ : \ {\cal D}({\cal A}) \ \longrightarrow\ L({\cal A})\ .
\end{equation}
For any state $\varphi \in {\cal D}({\cal A})$ its `orbit' of
$\cal A$ passing through $\varphi$ defines the Hilbert space
${\cal H}_\varphi$ with
\begin{equation}\label{}
    \< a\varphi|b\varphi\> = \varphi(a^*b)\ .
\end{equation}
Now, the embedding (\ref{e}) gives rise to the pull-backed bundle
$e^*(T^*{\cal A}^*)$. Its reduction by the left Gelfand ideal
$\mathcal{J}_\varphi$ at each point provides us with a GNS-bundle
which replaces the universal representation of a
$\mathbb{C}^*$-algebra (as a direct sum of all its irreducible
GNS-representations). When $\varphi$ is a pure state we obtain a
K\"ahlerian realization of a $\mathbb{C}^*$-algebra which
generalizes to the quantum setting the symplectic realization of a
Poisson manifold.

This bundle turns out to be related to the one defined by Shultz
\cite{Shultz}  (see also \cite{K-bundle}). We shall come back to
some of these bundle aspects in a forthcoming paper.

\section*{Acknowledgments}
A preliminary account of these results was presented in a series of
conferences: Holbaek Quantum Gravity Workshop (May 2008), MATHQCI
2008 CSIC Madrid (March 2008), XII Jornada SIMUMAT: {\it
Mathematical Structures of Quantum Mechanics}, {\it Geometry and
Quanta}, Toru\'n (June 2008). G.M. thanks the organizers of these
conferences for inviting him. D.C. thanks Beppe Marmo for the warm
hospitality in Naples where the main part of this paper was
prepared.


\end{document}